\documentclass[twocolumn]{revtex4-1}
\usepackage[utf8]{inputenc}
\usepackage{graphicx}
\usepackage{color}
\usepackage{amsmath}
\usepackage[left,running]{lineno}
\usepackage[utf8]{inputenc}
\usepackage{float}

\newcommand{\bea}{\begin{eqnarray}} 
\newcommand{\eea}{\end{eqnarray}}
\newcommand{\beq}{\begin{equation}}
\newcommand{\eeq}{\end{equation}}
\newcommand \hmu {\hat{\mu}}
\begin{document}
\title{Constraining the hadronic spectrum through QCD thermodynamics on the lattice}
\author{Paolo Alba$^a$, Rene Bellwied$^b$, Szabolcs Borsanyi$^{c}$, Zoltan Fodor$^{c,d,e}$, Jana G\"unther$^c$, Sandor D. Katz$^{d,f}$, Valentina Mantovani Sarti$^{g}$, Jacquelyn Noronha-Hostler$^b$, Paolo Parotto $^b$, Attila Pasztor$^{c}$, Israel Portillo Vazquez$^b$, Claudia Ratti$^{b}$}
\address{
$^b$ Department of Physics, University of Houston, Houston, TX 77204, USA\\
$^c$ Department of Physics, Wuppertal University, Gaussstr.  20, D-42119
Wuppertal, Germany\\
$^d$ Inst.  for Theoretical Physics, E\"otv\"os University, P\'azm\'any P. s\'et\'any 1/A, H-1117 Budapest, Hungary\\
$^e$ J\"ulich Supercomputing Centre, Forschungszentrum J\"ulich, D-52425 J\"ulich, Germany\\
$^f$ MTA-ELTE "Lend\"ulet" Lattice Gauge Theory Research Group,
P\'azm\'any P. s\'et\'any 1/A, H-1117 Budapest, Hungary\\
$^g$ Department of Physics, Torino University and INFN, Sezione di Torino, via P. Giuria 1, 10125 Torino, Italy}
\begin{abstract}
{
Fluctuations of conserved charges allow to study the chemical composition of
hadronic matter. A comparison between lattice simulations and the Hadron Resonance Gas (HRG)
model suggested the existence of missing strange resonances. To clarify this issue we
calculate the partial pressures of mesons and baryons with different strangeness
quantum numbers using lattice simulations in the confined phase of QCD. In order to make
this calculation feasible, we perform simulations at imaginary strangeness chemical potentials.
We systematically study the effect of different hadronic spectra on thermodynamic observables
in the HRG model and compare to lattice QCD results.
We show that, for each hadronic sector, the well established states are not
enough in order to have agreement with the lattice results. Additional states, either listed in the Particle Data Group booklet (PDG) but not well established, or predicted by the Quark Model (QM), are necessary in order to reproduce the lattice data.
For mesons, it appears that the PDG and the quark model do not list enough strange mesons,
or that, in this sector, interactions beyond those included in the HRG model are needed to 
reproduce the lattice QCD results.
}
\end{abstract}
\maketitle

\section*{Introduction}

The precision achieved by recent lattice simulations of QCD thermodynamics allows to extract, for the first time, quantitative predictions which provide a new insight into our understanding of strongly interacting matter. Recent examples include the precise determination of the QCD transition temperature \cite{Aoki:2006br,Aoki:2009sc,Borsanyi:2010bp,Bazavov:2011nk}, the QCD equation of state at zero \cite{Borsanyi:2010cj,Borsanyi:2013bia,Bazavov:2014pvz} and small chemical potential \cite{Borsanyi:2012cr,Gunther:2016vcp,Bazavov:2017dus} and fluctuations of quark flavors and/or conserved charges near the QCD transition \cite{Borsanyi:2011sw,Bazavov:2012jq,Bellwied:2015lba}. The latter are particularly interesting because they can be related to experimental measurements of particle multiplicity cumulants, thus allowing to extract the freeze-out parameters of heavy-ion collisions from first principles \cite{Karsch:2012wm, Bazavov:2012vg,Borsanyi:2013hza,Borsanyi:2014ewa,Noronha-Hostler:2016rpd}. Furthermore, they can be used to study the chemical composition of strongly interacting matter and identify the degrees of freedom which populate the system in the vicinity of the QCD phase transition \cite{Koch:2005vg,Bazavov:2013dta,Bellwied:2013cta}.

The vast majority of lattice results for QCD thermodynamics can be described, in the hadronic phase, by a non-interacting gas of hadrons and resonances which includes the measured hadronic spectrum up to a certain mass cut-off. This approach is commonly known as the Hadron Resonance Gas (HRG) model \cite{Dashen:1969ep,Venugopalan:1992hy,Karsch:2003vd,Karsch:2003zq,Tawfik:2004sw}. There is basically no free parameter in such a model, the only uncertainty being the number of states, which is determined by the spectrum listed in the Particle Data Book. It has been proposed recently to use the precise lattice QCD results on specific observables, and their possible discrepancy with the HRG model predictions, to infer the existence of higher mass states~\cite{NoronhaHostler:2008ju,Majumder:2010ik,Bazavov:2014xya}, not yet measured but predicted by Quark Model (QM) calculations \cite{Capstick:1986bm,Ebert:2009ub} and lattice QCD simulations \cite{Edwards:2012fx}. This leads to a better agreement between selected lattice QCD observables and the corresponding HRG curves. However, for other observables the agreement with the lattice gets worse, once the QM states are included.

\begin{figure}[ht!]
\centering
\includegraphics[width=0.48\textwidth]{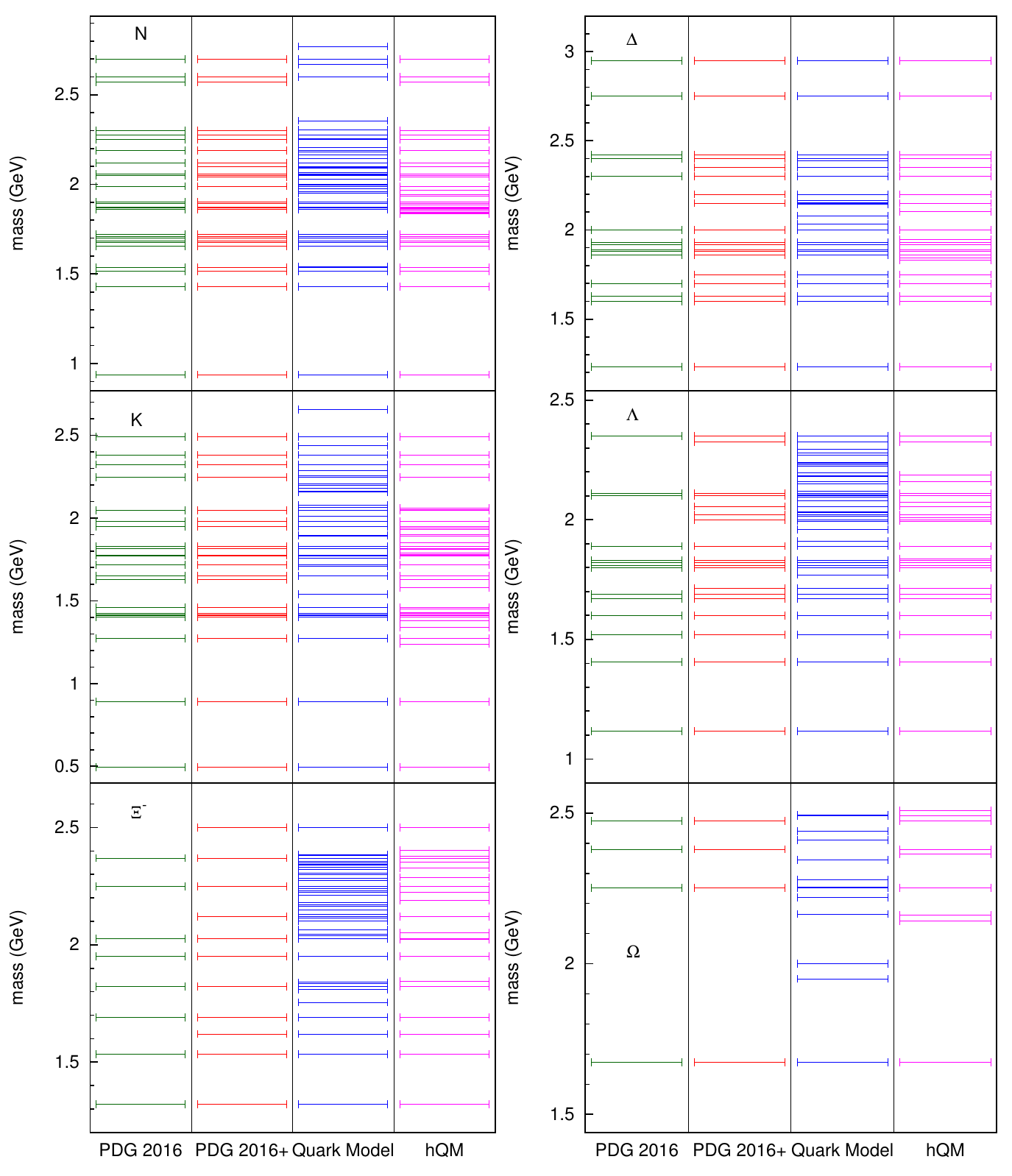}
\caption{\label{fig1}(Color online). Comparison of hadronic states, grouped according to the particle species, experimentally established in the PDG$2016$ (green), PDG$2016$ including one star states (red) \cite{Olive:2016xmw} and predicted by the QM (blue) \cite{Capstick:1986bm,Ebert:2009ub} and the hQM (magenta) \cite{Ferraris:1995ui,Ferretti:2015ada}.}
\end{figure} 
Amongst experimentally measured hadronic resonances within the Particle Data Group (PDG) list, there are different confidence levels on the existence of individual resonances.  The most well-established states are denoted by **** stars whereas * states indicate states with the least experimental confirmation. Furthermore, states with the fewest stars often do not have the full decay channel information known nor the branching ratios for different decay channels. 

In Fig.~\ref{fig1} we compare, for several particle species, the states listed in the PDG$2016$ (including states with two, three and four stars)~\cite{Olive:2016xmw}, in the PDG$2016+$ (including also states with one star)~\cite{Olive:2016xmw} and those predicted by the original Quark Model~\cite{Capstick:1986bm,Ebert:2009ub} and a more recent hypercentral version (hQM) \cite{Ferraris:1995ui}. The latter contains fewer states than the ones found in Refs. \cite{Capstick:1986bm,Ebert:2009ub}, due to inclusion of an interaction term between the quarks in the bound state, and the decay modes are listed for most of the predicted states. No mass cut-off has been imposed.
The total number of measured particles and anti-particles, excluding the charm and the bottom sector, increases from the $2016$ to the $2016+$ listing: considering particles and antiparticles and their isospin multiplicity we get 608 states with two, three and four stars and 738 states when we also include the one star states. In the QM description the overall increase is much larger: in total there are 1446 states in the non-relativistic QM \cite{Capstick:1986bm,Ebert:2009ub} and 1237 in the hQM \cite{Ferraris:1995ui,Ferretti:2015ada}. The QM predicts such a large number of states because they arise from all possible combinations of different quark-flavor, spin and momentum configurations. However, many of these states have not been observed in experiments so far; besides the basic QM description does not provide any information on the decay properties of such particles. As already mentioned, the hQM reduces the number of states by including an interaction term between quarks in a bound state. A more drastic reduction can be achieved by assuming a diquark structure \cite{Ferraris:1995ui,Santopinto:2004hw,Santopinto:2014opa} as part of the baryonic states, although experiments and lattice QCD may disfavor such a configuration \cite{Edwards:2012zza}.

\begin{figure}[ht!]
\includegraphics[width=0.48\textwidth]{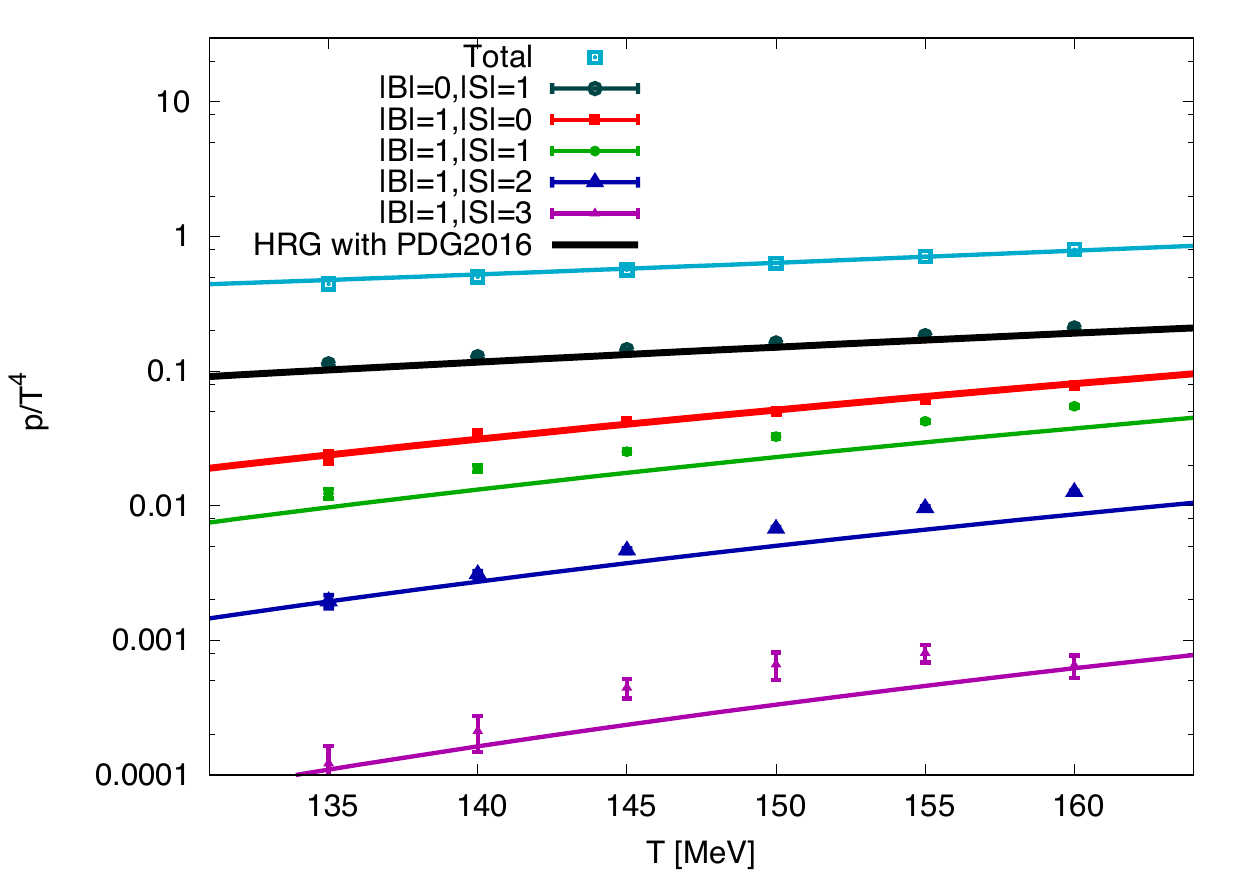}
\caption{\label{fig:p_log}(Color online) Logarithmic plot illustrating the many orders of magnitude the values of the partial pressures studied in this paper cover. The total pressure is taken from Ref. \cite{Borsanyi:2013bia}. Note, that the value for the $B=0$, $|S|=1$ 
sector is not a proper continuum limit, it is a continuum estimate based on the $N_t=12$ and $16$ lattices. For all other, the data are properly continuum extrapolated. In all cases, the solid lines correspond to the HRG model results based on the PDG2016 spectrum.}
\end{figure}

In this paper, we perform an analysis of several strangeness-related observables, by comparing the lattice QCD results to those of the HRG model based on different resonance spectra: the PDG 2016 including only the more established states (labeled with two, three and four stars), the PDG 2016 including all listed states (also the ones with one star), and the PDG 2016 with the inclusion of additional Quark Model states. This is done in order to systematically test the results for different 
particle species, and get differential information on the missing states, based on their strangeness content. The observables which allow the most striking conclusions are the partial pressures, namely the contribution to the total pressure of QCD from the hadrons, grouped according to their baryon number and strangeness content. 
The main result of this paper is a lattice determination of these partial pressures. This is a difficult task, since the partial pressures involve a cancellation of
positive and negative contributions (see the next section), and they span many orders of magnitude, as can be seen in Fig.~\ref{fig:p_log}.
From this analysis a consistent picture emerges: all observables confirm the need for not yet detected or at least not yet fully established strangeness states. The full PDG2016 list provides a satisfactory description of most observables, while for others the QM states are needed in order to reproduce the lattice QCD results. Besides, it appears that the PDG and the Quark Model do not list enough strange mesons or that, in this sector, interactions beyond those included in the HRG model are needed to reproduce the lattice QCD data \cite{Alba:2017bvm,Cabrera:2014lca}.

\section*{HRG and the strangeness sectors}

\begin{figure}[ht!]
\centering
\includegraphics[width=0.45\textwidth]{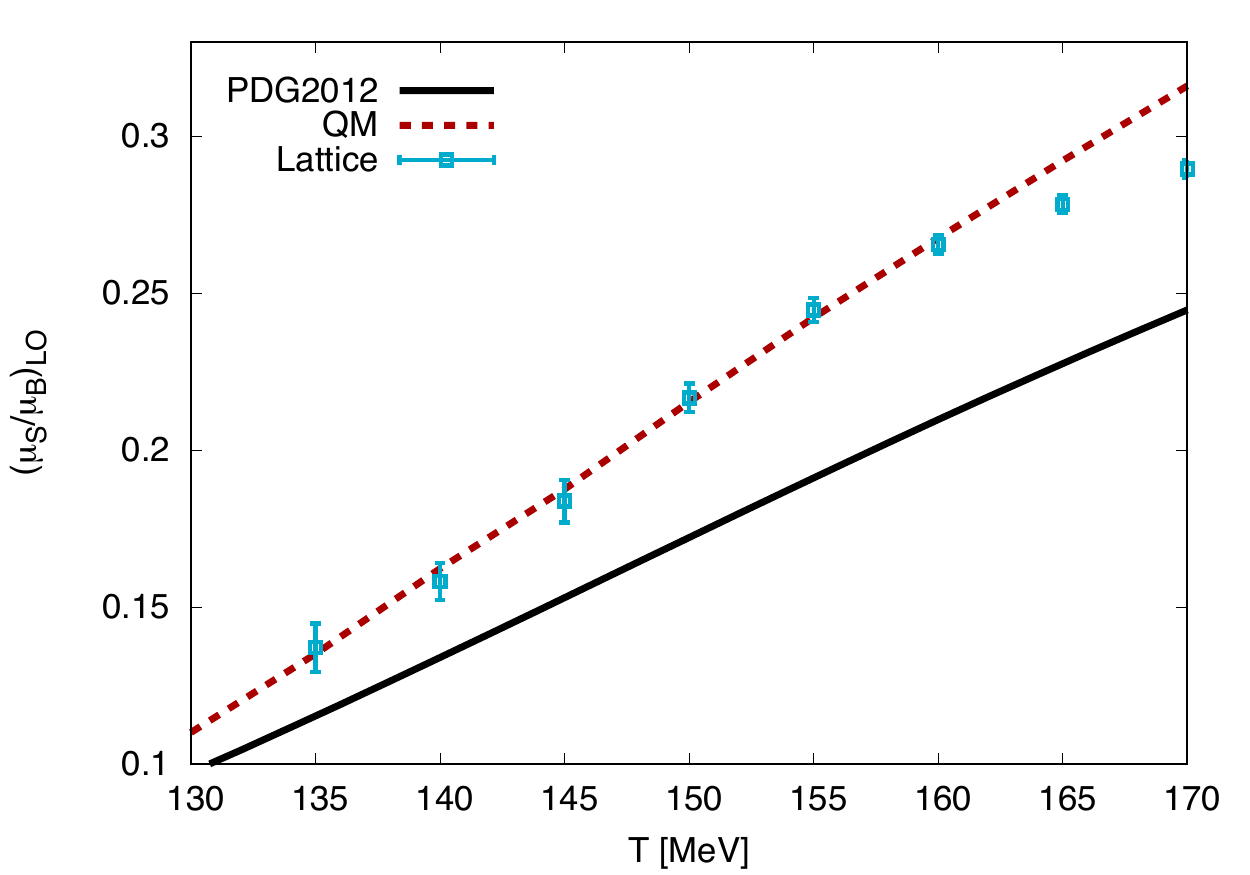}
\caption[]{\label{fig2}(Color online). Ratio $\mu_S/\mu_B$ at leading order as a function of the temperature. The HRG results are shown for different hadronic spectra, namely by using the PDG$2012$ (black solid line) and the QM (dashed red line).}
\end{figure}

\begin{figure}[t!]
\centering
\includegraphics[width=0.45\textwidth]{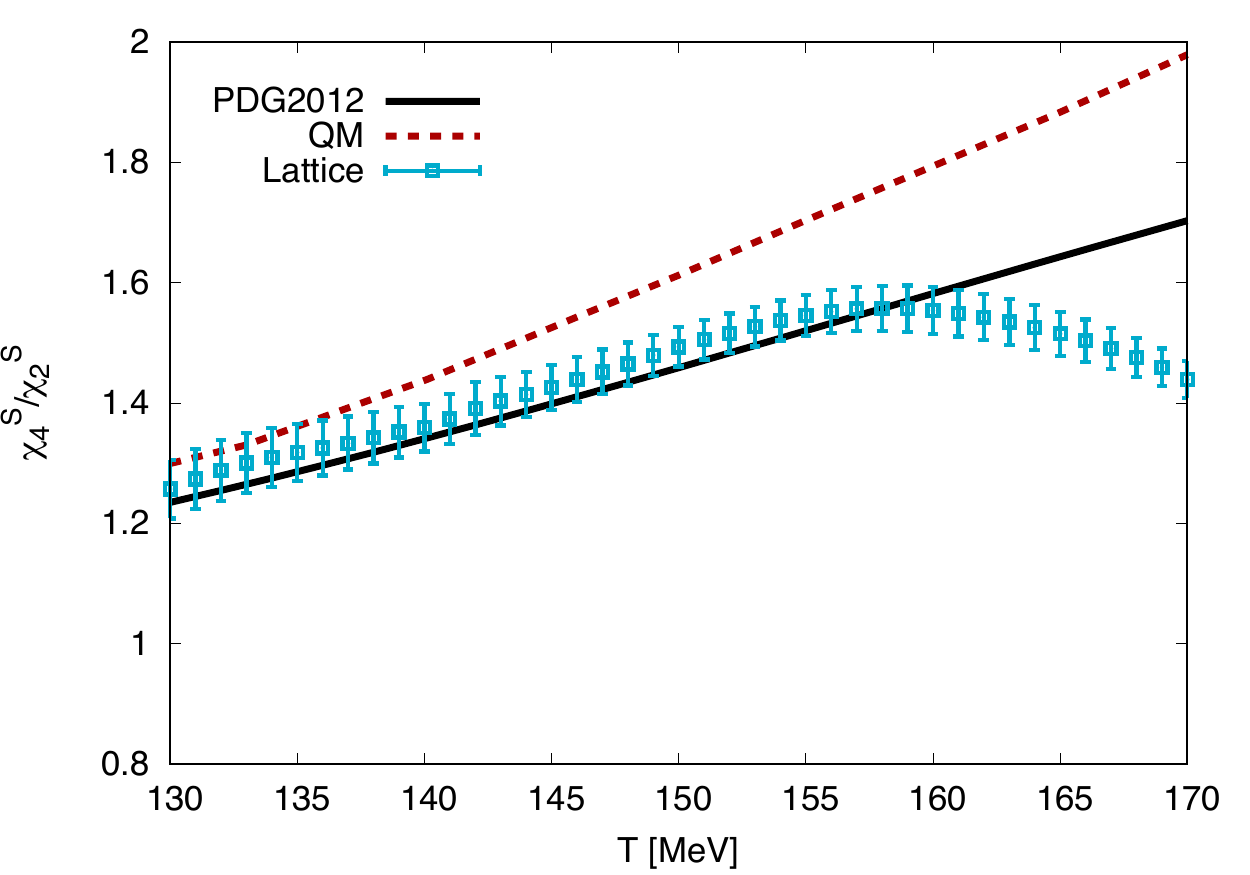}\\

\includegraphics[width=0.46\textwidth]{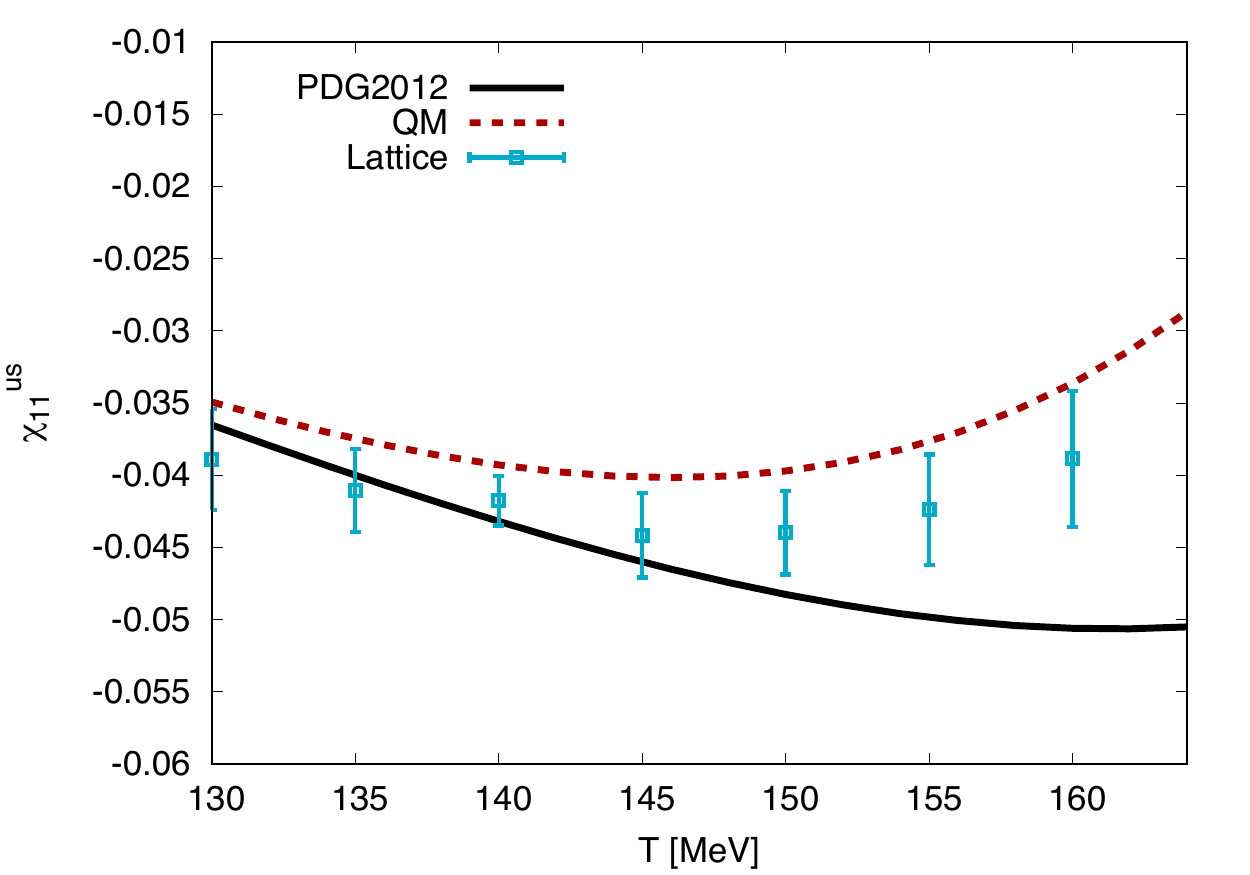}
\caption[]{\label{fig4}(Color online). Upper panel: Ratio $\chi_4 ^{S}/\chi_2 ^{S}$ as a function of the temperature. HRG model calculations based on the PDG$2012$ (black solid line) and the QM (red dashed line) spectra are shown in comparison to the lattice results from Ref.~\cite{Bellwied:2013cta}. Lower panel: comparison of up-strange correlator $\chi_{11} ^{us}$ simulated on the lattice~\cite{Bellwied:2015lba} and calculated in the HRG model using the PDG$2012$ (solid black line) and the QM (dashed red line) spectra.}
\end{figure}

The HRG model provides an accurate description of the thermodynamic properties of hadronic matter below $T_c$. This is especially true for global observables such as the total pressure and other collective thermodynamic quantities. However, it was recently noticed that more differential observables which are sensitive to the flavor content of the hadrons show a discrepancy between HRG model and lattice results \cite{Bazavov:2014xya}. An example of such discrepancy is shown in Fig. \ref{fig2} and will be explained below. Such observables involve the evaluation of susceptibilities of conserved charges in the system at vanishing chemical potential:
\begin{equation}
\chi_{lmn} ^{BQS} =\left( \dfrac{\partial ^{l+m+n}P(T,\mu_B,\mu_Q,\mu_S)/T^4}{\partial
(\mu_B/T)^l \partial (\mu_Q/T)^m \partial (\mu_S/T)^n}\right)_{\mu=0}.
\end{equation}\label{eq2}
Cumulants of net-strangeness fluctuations and correlations with net-baryon number and net-electric charge have been evaluated on the lattice in a system of $(2+1)$ flavours at physical quark masses and in the continuum limit~\cite{Borsanyi:2014tda,Borsanyi:2013hza,Bellwied:2015lba}.
The same quantities can be obtained within the HRG model. 
In this approach,
the total pressure in the thermodynamic limit for a gas of non-interacting particles in the grand-canonical ensemble is given by:
\begin{eqnarray}
&&P_{tot}(T,{\mu}) = \sum _k P_k(T,\mu_k)
= \sum _k (-1)^{B_k+1} \dfrac{d_k T}{(2\pi) ^3}  \int d^3\vec{p} \nonumber\\
&&   \ln  \left(1+ (-1)^{B_k+1} \exp \left[ -\dfrac{(\sqrt{\vec{p} ^2+m_k^2}-\mu_k)}{T} \right] \right) \rm{,}
\end{eqnarray}\label{eq1}
where the sum runs over all the hadrons and resonances included in the model. Here the single particle chemical potential is defined with respect to the global conserved charges (baryonic $B$, electric $Q$ and strangeness $S$) as $\mu_k=B_k\mu_B+Q_k\mu_Q+S_k\mu_S$.
More details on the HRG model used here can be found in Ref.~\cite{Alba:2014eba}. In order to describe the initial conditions of the system occurring during a heavy-ion collision, we require strangeness neutrality and the proper ratio of protons to baryons given by the colliding nuclei, $n_Q=\tfrac{Z}{A} n_B\simeq0.4n_B$.
These conditions yield $\mu_S$ and $\mu_Q$ as functions of $\mu_B$; their specific dependence on $\mu_B$ is affected by  the amount of strange particles and charged particles included in the model. 
To leading order in $\mu_B$, the ratio $\mu_S/\mu_B$ reads~\cite{Borsanyi:2013hza, Bazavov:2012vg}:
\begin{equation}
\left( \dfrac{\mu_S}{\mu_B}\right) _{LO}=-\dfrac{\chi_{11} ^{BS}}{\chi_2 ^{S}}-\dfrac{\chi_{11} ^{QS}}{\chi_2^{S}}\dfrac{\mu_Q}{\mu_B} \rm{.}
\end{equation}\label{eq3}
The inclusion of a larger number of heavy hyperons, such as $\Lambda$ and $\Xi$, and the constraint of strangeness neutrality are reflected by a larger value of the strange chemical potential $\mu_S$ as a function of temperature and baryo-chemical potential. 
In Fig.~\ref{fig2} this ratio is shown as a function of the temperature: our new, continuum extrapolated lattice results are compared to the HRG model calculations based on the 2012 version of the PDG and on the Quark Model states (as done in Ref. \cite{Bazavov:2014xya}). One should expect agreement between HRG model and lattice calculations up to the transition temperature which has been determined independently on the lattice to be $\sim155$ MeV \cite{Aoki:2006br,Aoki:2009sc,Borsanyi:2010bp,Bazavov:2011nk}. The HRG model based on the QM particle list yields a better agreement with the lattice data within error bars, while the HRG results based on the PDG2012 spectrum underestimate the data. However, for other observables such as $\chi_4^S/\chi_2^S$ and $\chi_{11}^{us}$ (see the two panels of Fig. \ref{fig4}), the agreement between HRG model and lattice results is spoiled when including the QM states. The QM result overestimates both $\chi_4^S/\chi_2^S$ and $\chi_{11}^{us}$; $\chi_4^S/\chi_2^S$ is proportional to the average strangeness squared in the system: the fact that the QM overestimates it, means that it contains either too many multi-strange states or not enough $|S|=1$ states. Moreover, the contribution to $\chi_{11}^{us}$ is positive for baryons and negative for mesons: this observable provides the additional information that the QM list contains too many (multi-)strange baryons or not enough $|S|=1$ mesons.

In this paper, we try to solve this ambiguity, even though we are aware that it might be difficult to resolve the contribution of high mass particles in our simulations. We separate the pressure of QCD as a function of the temperature into contributions coming from hadrons grouped according to their quantum numbers. This is done by assuming that, in the low temperature region we are interested in, the HRG model in the Boltzmann approximation yields a valid description of QCD thermodynamics. If this is the case, the pressure of the system can be written as~\cite{Bazavov:2013dta,Noronha-Hostler:2016rpd}:
\begin{eqnarray}
\label{eq:pressure}
P(\hmu_B,\hmu_S) &=& P^{BS}_{00}+P^{BS}_{10}\cosh(\hmu_B)+P^{BS}_{01} \cosh(\hmu_S) 
\nonumber \\
&+& P^{BS}_{11} \cosh(\hmu_B-\hmu_S)
\nonumber \\
&+& P^{BS}_{12} \cosh(\hmu_B-2\hmu_S)
\nonumber \\
&+& P^{BS}_{13} \cosh(\hmu_B-3\hmu_S) 
\;,
\end{eqnarray}
where $\hmu_i=\mu_i/T$, and the quantum numbers can be understood as absolute values. These partial pressures are
the main observables we study. Notice that we do not distinguish the particles according to their electric charge content.

Assuming this ansatz for the 
pressure, the partial pressures $P^{BS}_{ij}$ can be expressed as linear combinations of
the susceptibilities $\chi^{BS}_{ij}$. An example of one such formula is:
\begin{equation}
\label{eq:Smeson}
P^{BS}_{01} = \chi^{S}_{2} - \chi^{BS}_{22} \rm{,}
\end{equation}
which gives the strange meson contribution to the pressure. 
This means that in principle one could determine these partial pressures 
directly from $\mu=0$ simulations, by evaluating linear combinations 
of the $\chi^{BS}_{ij}$ directly. This can be done on the lattice, by
calculating fermion matrix traces, that can be evaluated with the help
of random sources~\cite{Allton:2002zi,Bellwied:2015lba}.

This is not the approach we pursue here, since the noise level in the calculation would be too high, certainly for the $S=2$ or $3$ sectors, but as Fig.~\ref{fig:lattice}(bottom) 
shows, probably already for $S=1$. The higher order fluctuations
are already quite noisy, because they involve big cancellations between 
positive and negative contributions~\cite{Bellwied:2015lba}. In addition, when we 
take linear combinations to calculate the partial 
pressures, we introduce
extra cancellations between the susceptibilities. Therefore we propose to use
an imaginary strangeness chemical potential and extract the partial pressures from the
$\operatorname{Im} \mu_S$ depdendence of low order susceptibilities. For earlier works
exploiting imaginary chemical potentials, see \cite{deForcrand:2002hgr, DElia:2002tig, Wu:2006su, DElia:2007bkz, Conradi:2007be, deForcrand:2008vr, DElia:2009pdy}. A more
recent work, that uses imaginary chemical potentials to estimate higher order 
susceptibilities is \cite{DElia:2016jqh}.

\begin{figure}[t!]
\vspace{-4.5cm}
\centering
\includegraphics[trim={0 2.5cm 0 0}, width=0.72\textwidth]{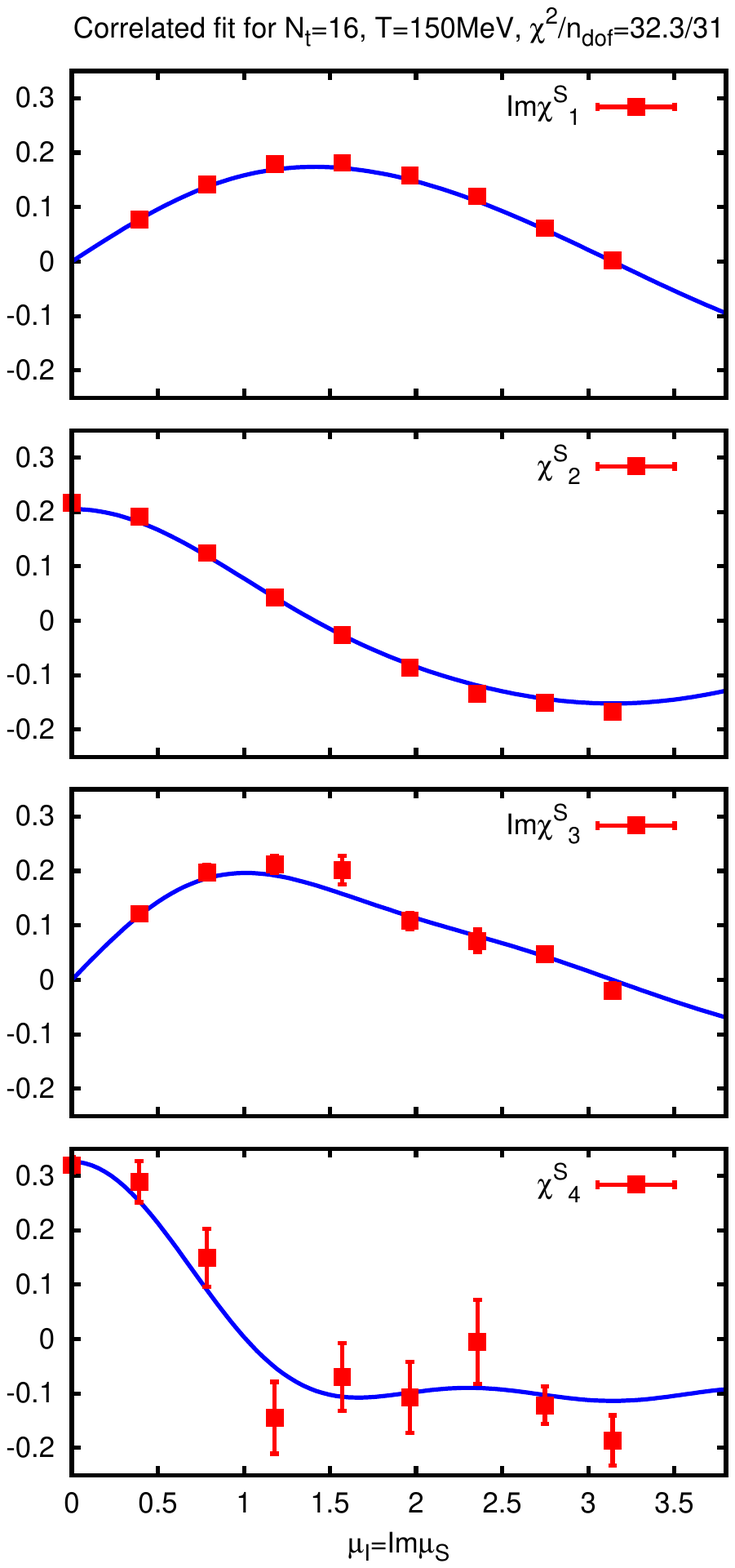}
\caption{
\label{fig:corrfit}(Color online) An example of a correlated fit for the quantities $\chi^S_1$, $\chi^S_2$, $\chi^S_3$ and $\chi^S_4$.}
\end{figure}

\begin{figure}[t!]
\centering
\includegraphics[trim={0 2cm 0 2cm},clip,width=0.53\textwidth]{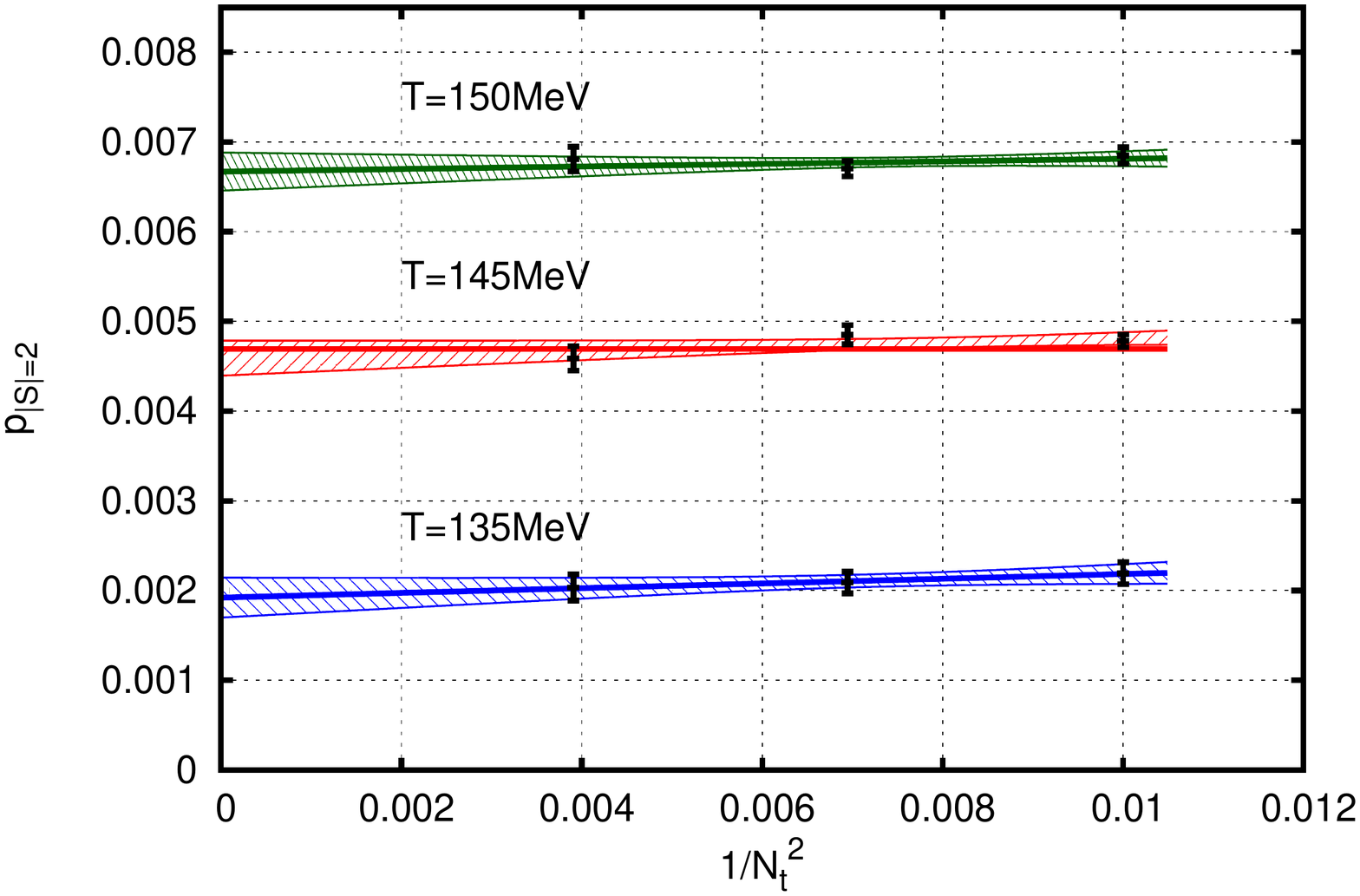}
\hspace{-7cm}
\includegraphics[trim={0 1.5cm 0 2cm},clip,width=0.52\textwidth]{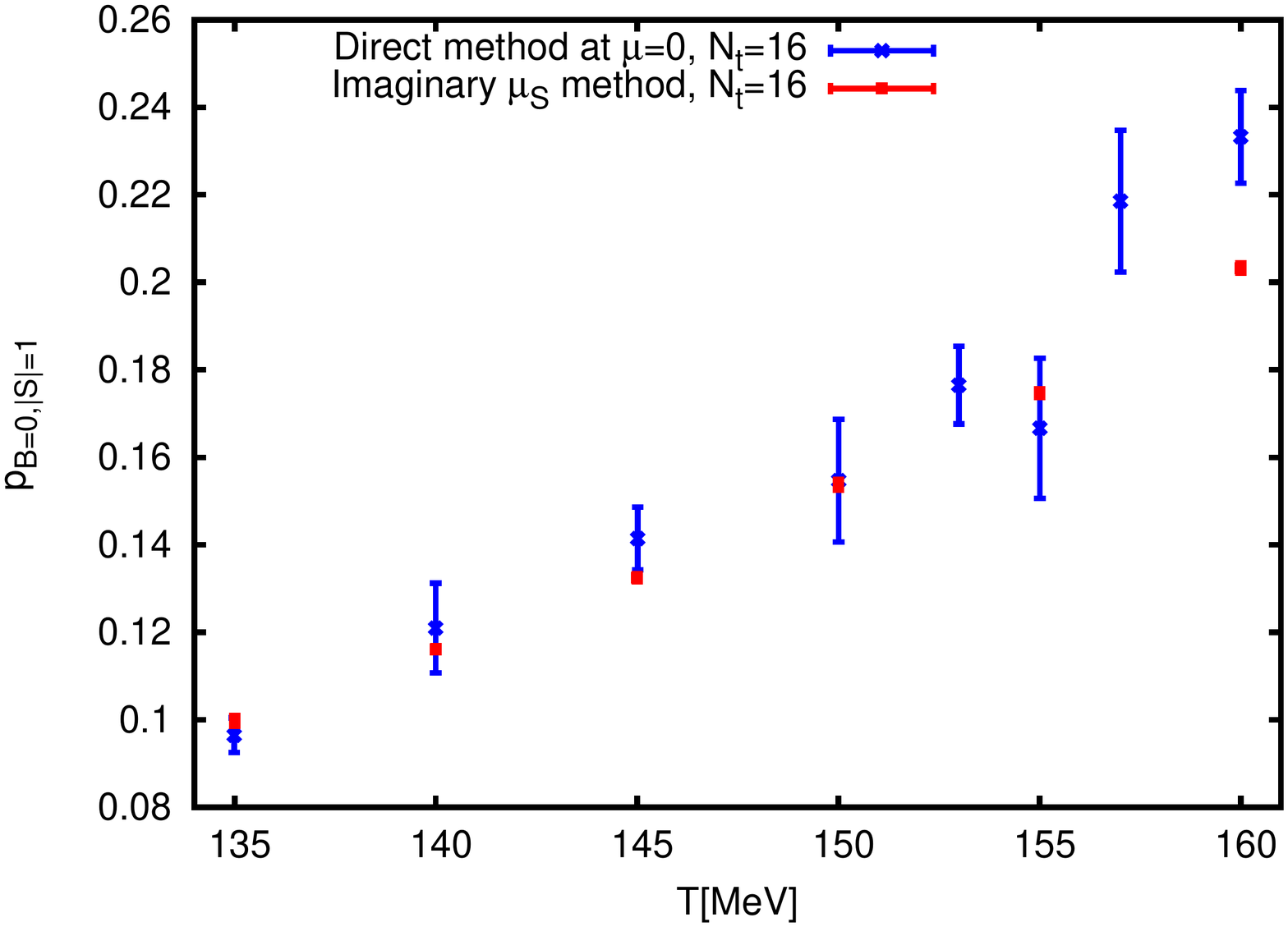}
\caption{\label{fig:lattice}(Color online) Upper panel: Examples of continuum limit extrapolation 
from our $N_t=10,12,16$ lattices. Lower panel: Comparison of our method to Taylor 
expansion from $\mu=0$ data for $P_{01}$. The statistics would
explain only a factor of 2 difference in the errorbars, but the 
improvement is much more drastic than that.}
\end{figure}

\begin{figure}[ht!]
\includegraphics[width=0.48\textwidth]{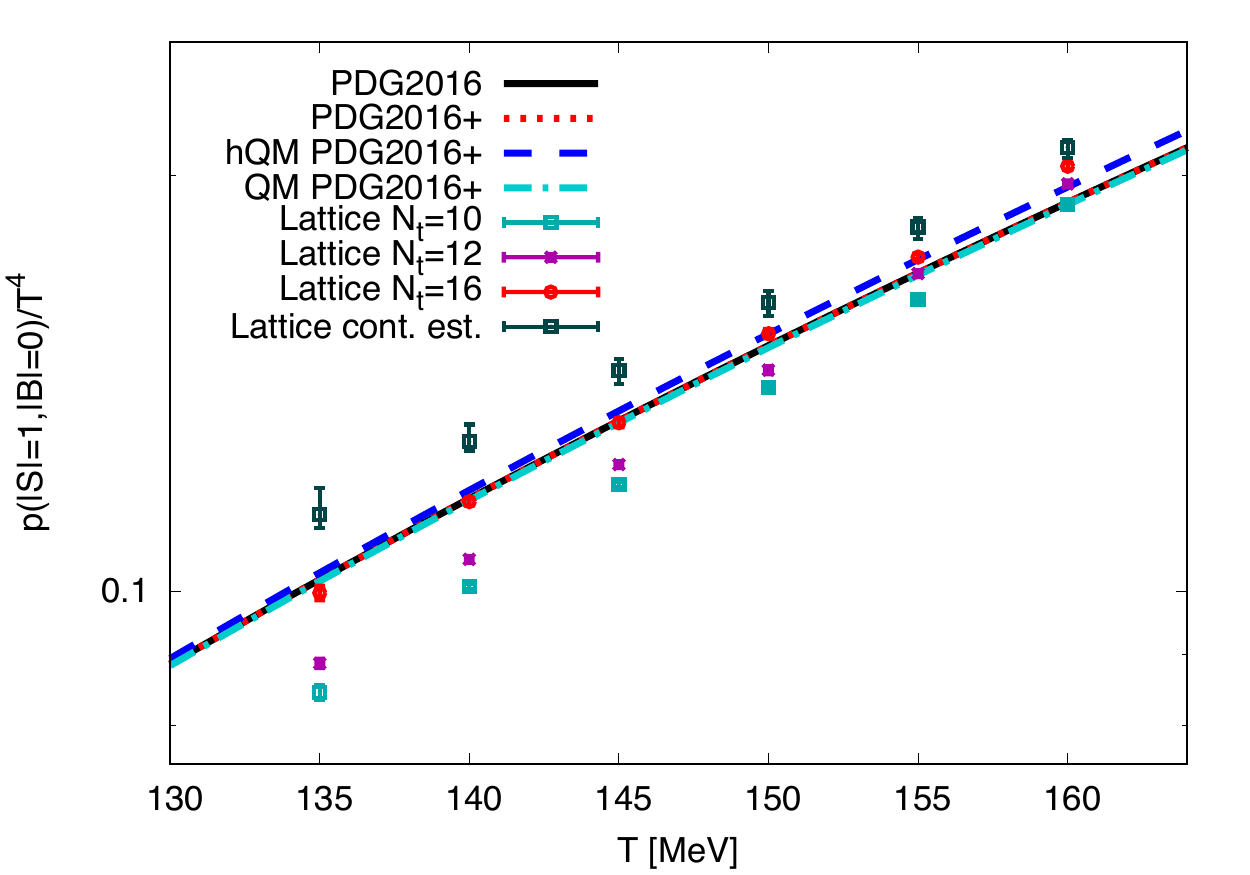}
\caption{\label{fig:strangemeson}(Color online).
Comparison between the lattice results for the partial pressures of strange mesons and the HRG model predictions. 
No continuum extrapolation could be carried out from our $N_t=10,12,16$ data. The plot includes the lattice
data at finite lattice spacings for $N_t=10,12,16$ and a continuum estimate form the $N_t=12$ and $16$ lattices. The points are the lattice results, while the curves are  PDG$2016$ (solid black), PDG$2016+$ (including one star states, red dotted), PDG$2016+$ and additional states from the hQM (blue, dashed) \cite{Ferraris:1995ui,Ferretti:2015ada}, PDG$2016+$ and additional states from the QM (cyan, dash-dotted) \cite{Capstick:1986bm,Ebert:2009ub}.
}
\end{figure}
\section*{Lattice method}
\begin{figure*}[ht]
\centering
\includegraphics[width=0.49\textwidth]{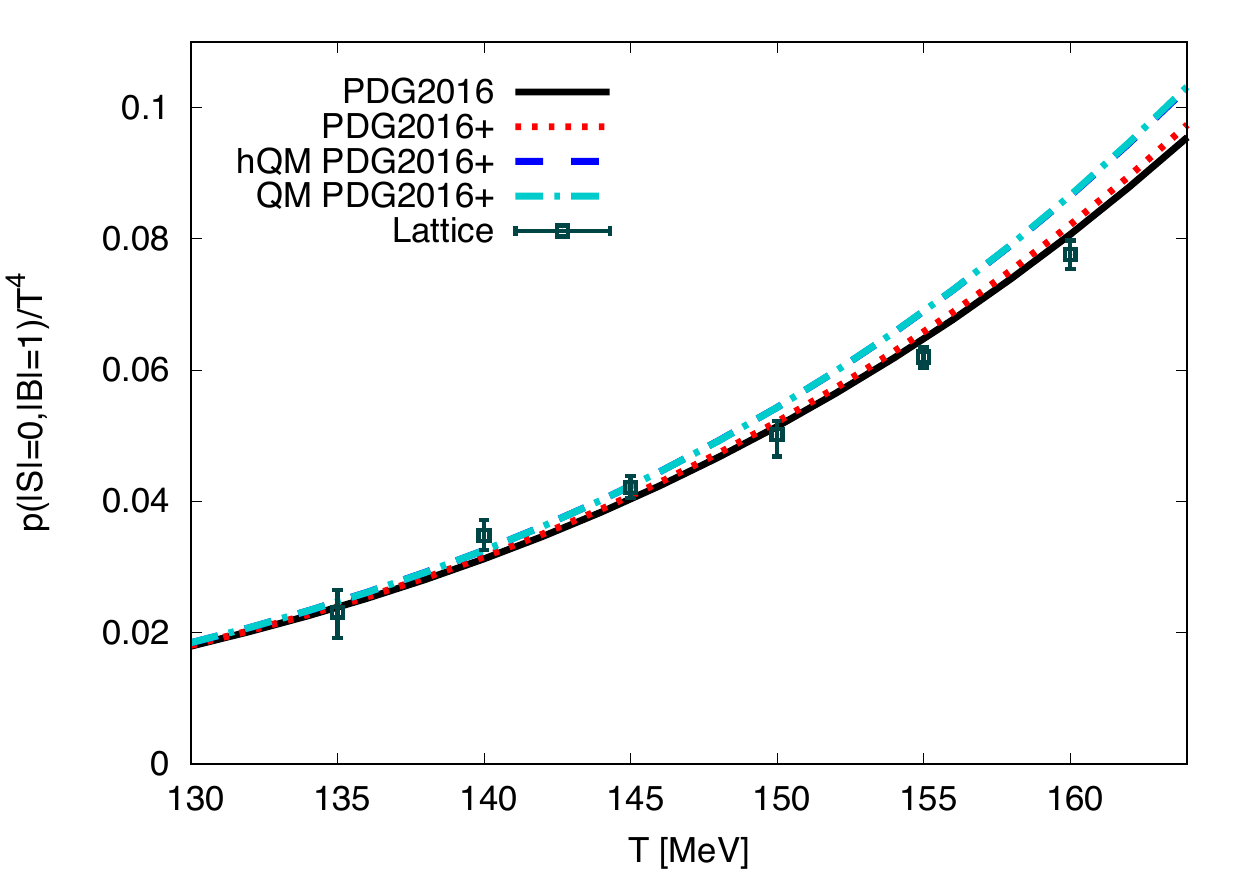}
\includegraphics[width=0.49\textwidth]{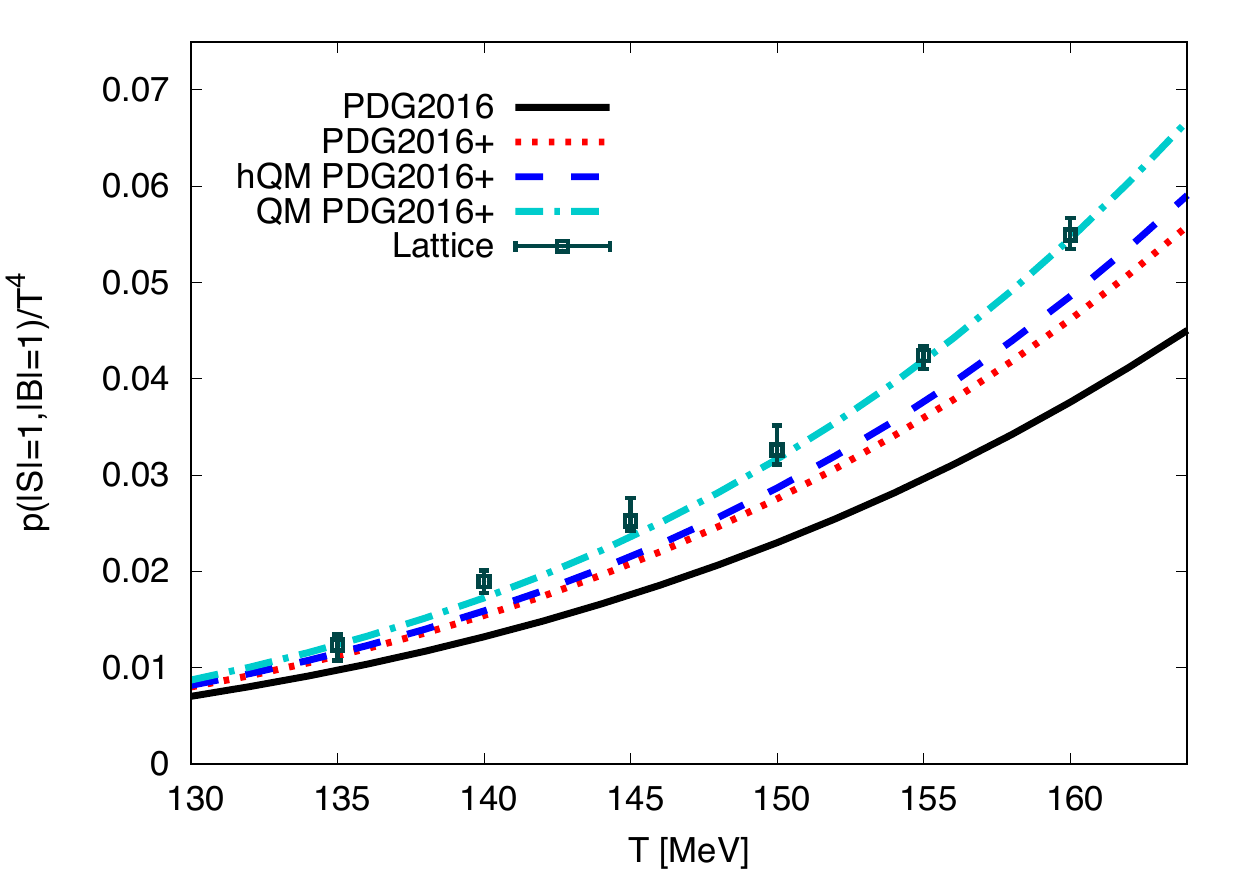}
\includegraphics[width=0.49\textwidth]{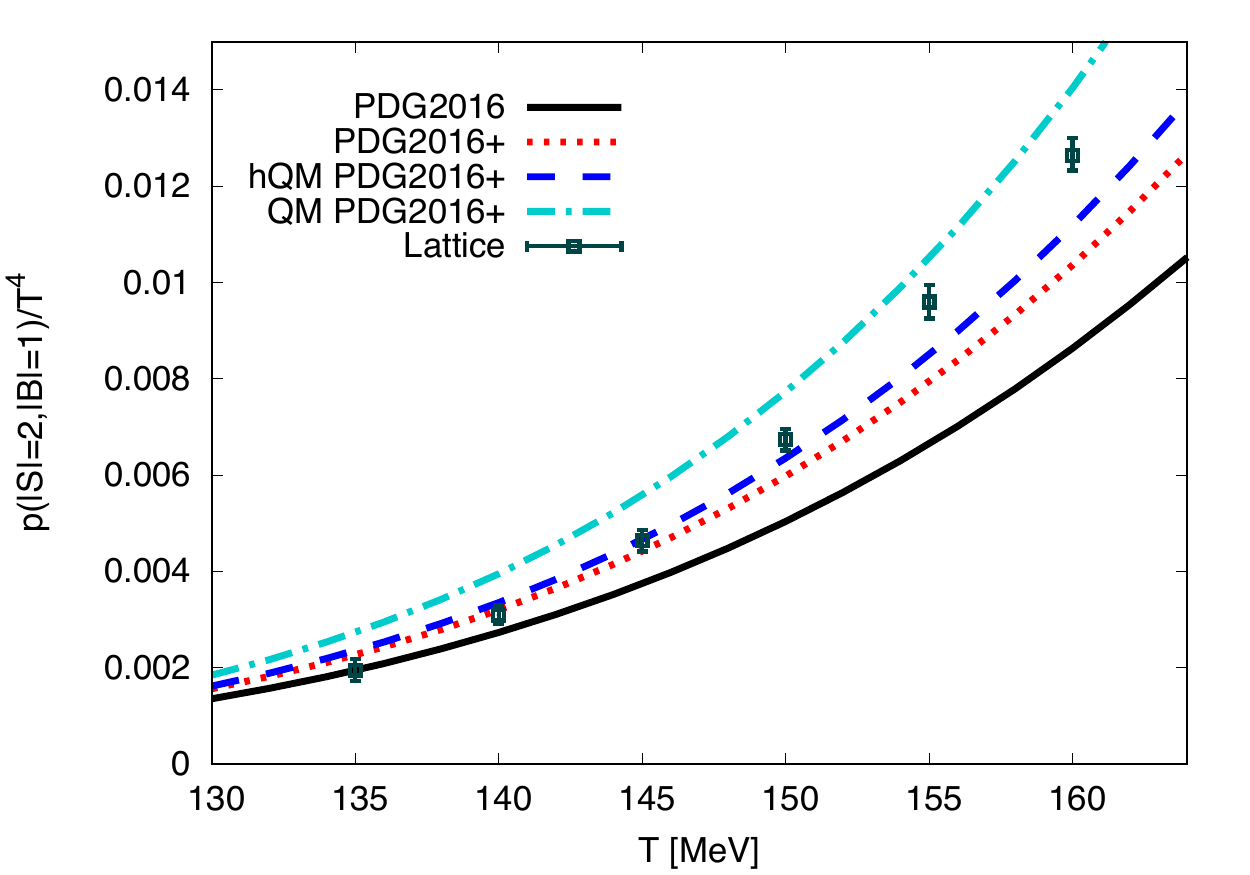}
\includegraphics[width=0.49\textwidth]{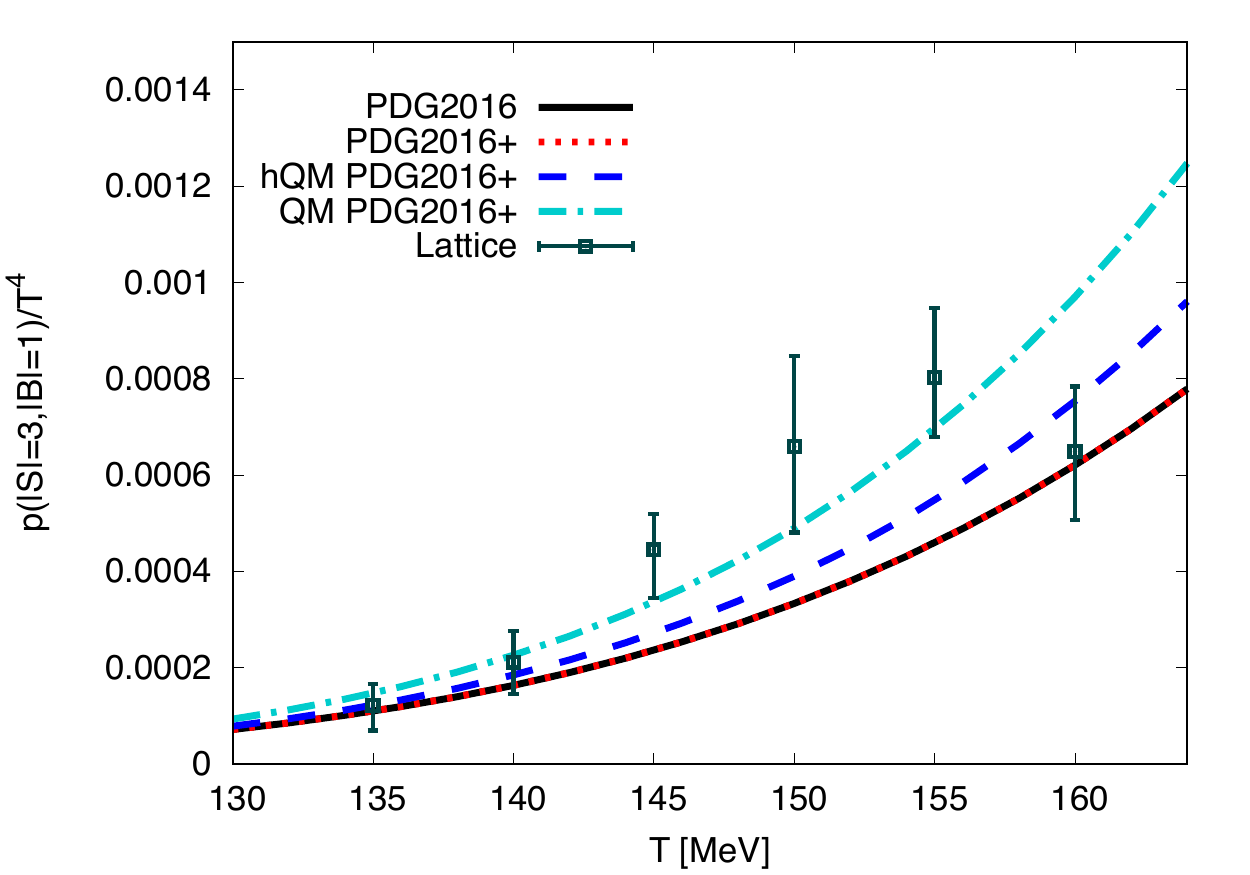}
\caption{\label{fig5}(Color online). Comparison between the lattice results for the partial pressures and the HRG model predictions. Upper panels: non-strange baryons (left), $|S|=1$ baryons (right). Lower panels: $|S|=2$ baryons (left), $|S|=3$ baryons (right). The points are the lattice results, while the curves are  PDG$2016$ (solid black), PDG$2016+$ (including one star states, red dotted), PDG$2016+$ and additional states from the hQM (blue, dashed) \cite{Ferraris:1995ui,Ferretti:2015ada}, PDG$2016+$ and additional states from the QM (cyan, dash-dotted) \cite{Capstick:1986bm,Ebert:2009ub}.}
\end{figure*}
Our lattice simulations use the same 4stout staggered action
as ~\cite{Bellwied:2015lba,Bellwied:2015rza,Borsanyi:2016ksw,Gunther:2016vcp}. 
We generate configurations at $\mu_B=\mu_S=\mu_Q=0$ as well as $\mu_B=\mu_Q=0$ and 
$\operatorname{Im} \mu_S>0$, in the temperature range $135 \rm{MeV} \leq T \leq 165 \rm{MeV}$.
All of our lattices have an aspect ratio of $LT=4$. We run roughly 
$1000-2000$ configurations at each simulation point, separated by 10 HMC trajectories.
For the determination of the strangeness sectors we use the HRG ansatz of 
equation~(\ref{eq:pressure}) for the pressure.
With the notation $\mu_S= i \mu_I$ we get by simple differentiation:
\begin{align}
\operatorname{Im}\chi^B_1 =         - P^{BS}_{11}\sin(\mu_I)         -  P^{BS}_{12}\sin(2\mu_I) \nonumber \\
                                                           - P^{BS}_{13} \sin(3\mu_I) \rm{,} \nonumber \\
\chi^B_2                  =  P^{BS}_{10} \ + P^{BS}_{11}\cos(\mu_I)       +  P^{BS}_{12}\cos(2\mu_I) \nonumber \\  
                                                                  +  P^{BS}_{13} \cos(3\mu_I) \rm{,}           \\
\operatorname{Im}\chi^S_1 =          (P^{BS}_{01}+P^{BS}_{11})\sin(\mu_I) + 2 P^{BS}_{12}\sin(2\mu_I)        \nonumber \\
                                                                 + 3 P^{BS}_{13} \sin(3\mu_I) \rm{,} \nonumber
\end{align}
and similar terms for $\chi^S_2$, $\chi^S_3$ and $\chi^S_4$. An advantage of the imaginary chemical potential approach is that we do not have to make any extra assumptions, as compared to the direct evaluation of the
linear combinations, see equation~(\ref{eq:Smeson}), as the linear combinations there were
already derived from the HRG ansatz. On the other hand, our method reduces the errors considerably,
as the lower derivatives already contain the information on the higher strangeness sectors.
Simulations at imaginary chemical potential are not hampered by the sign problem, so the 
evaluation of the lower order susceptibilities at $\operatorname{Im} \mu_S>0$ is not any harder than at $\mu=0$.

To obtain the Fourier coefficients $(P^{BS}_{01}+P^{BS}_{11})$, $P^{BS}_{12}$ and $P^{BS}_{13}$ we 
perform a correlated fit with the previous ansatz for the observables 
$\chi^S_1$, $\chi^S_2$, $\chi^S_3$ and $\chi^S_4$ at every temperature. To obtain $P^{BS}_{10}$ we fit
$\chi^B_2$, to obtain $P^{BS}_{11}$ we fit $\chi^B_1$. We note that $P^{BS}_{11}$ could be deduced
from $\chi^B_2$ as well, but with considerably higher statistical errors.
To get $P^{BS}_{01}$ we just take 
the difference $(P^{BS}_{01}+P^{BS}_{11})-P^{BS}_{11}$. 
As an illustration that the HRG based ansatz fits our lattice data we include an example of
the correlated fit for $\chi^S_1$, $\chi^S_2$, $\chi^S_3$ and $\chi^S_4$ in Fig.~\ref{fig:corrfit}

For statistical errors, we use the jackknife
method. For the continuum limit we use $N_t=10,12$ and $16$ lattices. 
To estimate the systematic errors we repeat the anaylsis in several different ways: to connect the 
lattice parameters to physical temperatures we use two different scale settings, based on
 $w_0$ and $f_\pi$. More details on the scale setting
can be found in~\cite{Bellwied:2015lba}. 
For each choice of the scale settings we use two different spline interpolations for the temperature dependence of the $P^{BS}_{ij}$. Both of these describe the data well. For each of these four choices we do the continuum limit in four 
different ways, by applying tree level improvement or not, and by using a straight $a+b/N_t^2$ and 
a rational function $1/(a+b/N_t^2)$ ansatz for the continuum limit. These 16 different results are then weighted with the AKAIKE information 
criterion~\cite{akaike1974new} using the histogram method~\cite{Durr:2008zz}.
Examples of linear continuum limit extrapolations are included in Fig.~\ref{fig:lattice} (top). 
For the $B=0$, $S=1$ sector, which includes a large contribution 
from kaons, the continuum extrapolation could not be carried out using these lattices. 
For this case we obtain a continuum estimate, based on the assumption that only $N_t=10$ is
not in the scaling regime, therefore using only the $N_t=12$ and $N_t=16$ 
lattices, and the same sources of systematic error as before, but now with uniform weights.

Finally, as a comparison we show in Fig.~\ref{fig:lattice} (bottom) 
one of the partial pressures determined with both methods for $N_t=16$, 
using the same action. The figure shows that using imaginary chemical 
potential improved the accuracy drastically already in the $S=1$ sector. In the $S=2,3$ sectors, 
the direct method would be too noisy to plot, while the imaginary $\mu$ method allows for a quite 
accurate determination of the strangeness sectors.

\begin{figure}[ht]
\includegraphics[width=0.44\textwidth]{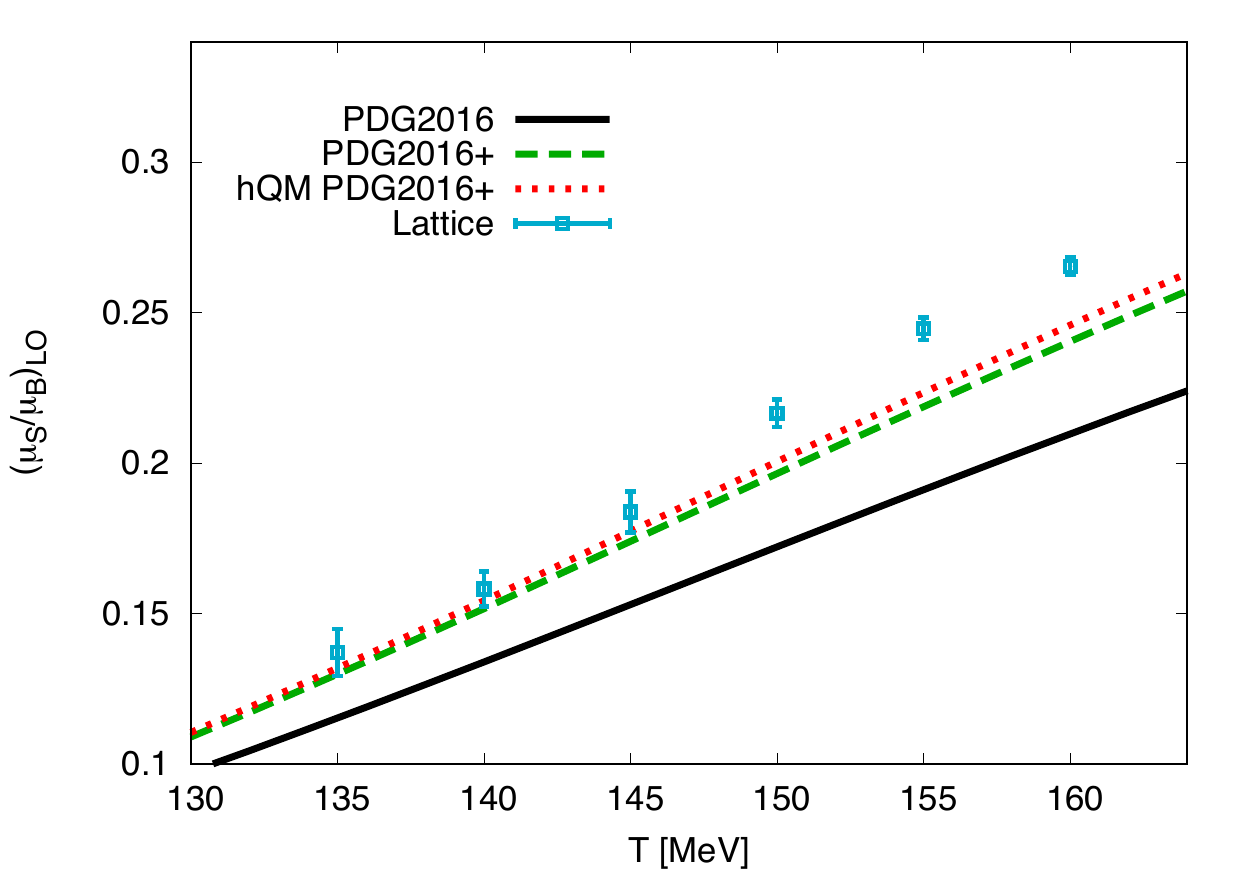}

\includegraphics[width=0.45\textwidth]{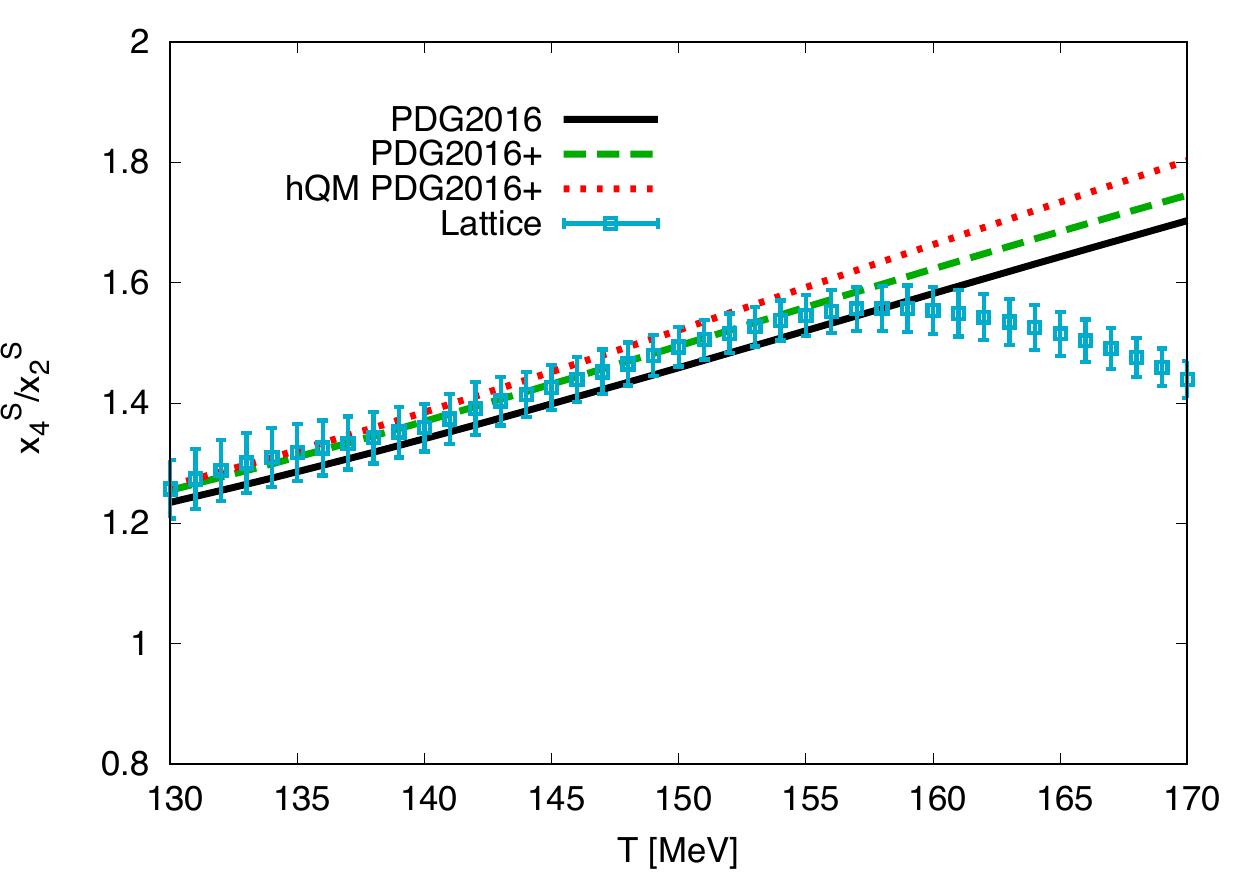}
\caption[]{\label{fig6}(Color online). Upper panel: Ratio $(\mu_S/\mu_B)_{LO}$ as a function of the temperature. Lower panel: $\chi_4 ^{S}/\chi_2 ^{S}$ as a function of the temperature. In both cases, the lattice results are compared to the HRG model curves based on the PDG 2016 (black, solid line), the PDG2016+ (green, dashed line) and the PDG2016+ with additional states from the hQM (red, dotted line).  
}
\end{figure}

\section*{Results and their interpretation}
We evaluate the contributions to the total QCD pressure from the following sectors: strange mesons, non-strange baryons, and baryons with $|S|=1,~2,~3$. For each sector, we compare the lattice QCD results to the predictions of the HRG model using the PDG2016, PDG2016+, hQM and QM spectra.

In Figs.~\ref{fig:strangemeson} and \ref{fig5} we show our results. 
Fig.~\ref{fig:strangemeson} shows the contributrion of strange mesons, while Fig.~\ref{fig5} shows the contribution of non-strange baryons (upper left), $|S|=1$ baryons (upper right),
$|S|=2$ baryons (lower left) and $|S|=3$ baryons (lower right).

We observe that, in all cases except the non-strange baryons, the established states from the most updated version of the PDG are not sufficient to describe the lattice data. For the baryons with $|S|=2$, a considerable improvement is achieved when the one star states from PDG2016 are included. The inclusion of the hQM states pushes the agreement with the lattice results to higher temperatures, but one has to keep in mind that the crossover nature of the QCD phase transition implies the presence of quark degrees of freedom in the system above $T\simeq155$ MeV, which naturally yields a deviation from the HRG model curves. Notice that, in the case of $|S|=1,~3$ baryons, it looks like even more states than PDG2016+ with hQM are needed in order to reproduce the lattice results: the agreement improves when the resonances predicted by the QM \cite{Capstick:1986bm,Ebert:2009ub} are added to the spectrum. 
Fig.~\ref{fig:p_log} shows the relative contribution of the sectors to the total pressure. Notice that three orders of magnitude separate the $|S|$=1 meson contribution from the $|S|=3$ baryon one. The method we used for this analysis, namely simulations at imaginary $\mu_S$, was crucial in order to extract a signal for the multi-strange baryons.

As for strange mesons, we point out that the PDG2016 and 2016+ coincide since there is no star ranking for mesons. In this sector, it was not possible to perform a continuum extrapolation for the data, since apparently they are not in the scaling regime. However, there is clear trend in the $N_t=10,~12,~16$ data that makes it very natural to assume that the continuum extrapolated results will lie above the HRG curves.
We also include a continuum estimate of this quantity, based on only the $N_t=12$ and $16$ lattices, which is clearly above the HRG curves.
This might mean that, for strange mesons, the interaction between particles is not well mimicked by the HRG model in the Boltzmann approximation, or that we need even more states than the ones predicted by the QM. This was already suggested in Ref.~\cite{Lo:2015cca}, based on a different analysis. In general, one should keep in mind that here we use a version of the HRG model in which particles are considered stable (no width is included). Any width effects on the partial pressures can be considered in future work. Analogously, our previous lattice QCD results did not show indications of finite volume effects for the total pressure. These effects have not been checked for the partial pressures presented here. 

Our analysis shows that, for most hadronic sectors, the spectrum PDG2016, does not yield a satisfactory description of the lattice results. All sectors clearly indicate the need for more states, in some cases up to those predicted by the original Quark Model. One has to keep in mind that using the QM states in a HRG description will introduce additional difficulties in calculations used in heavy ion phenomenology, as the QM does not give us the decay properties of these new states.
The HRG model is successfully used to describe the freeze-out of a heavy-ion collision, by fitting the yields of particles produced in the collision and thus extracting the freeze-out temperature and chemical potential \cite{Cleymans:2005xv,Manninen:2008mg,Andronic:2011yq}, which are known as ``thermal fits". To this purpose, one needs to know the decay modes of the resonances into the ground state particles which are reaching the detector. As of yet, the QM decay channels are unknown so predictions for their decay channels are needed first, before one can use them in thermal fits models.

In conclusion, we re-calculate the two observables which triggered our analysis, namely $(\mu_S/\mu_B)_{LO}$ and $\chi_4^S/\chi_2^S$, with the updated hadronic spectra. They are shown in the two panels of Fig. \ref{fig6}. The upper panel shows $(\mu_S/\mu_B)_{LO}$ as a function of the temperature: the lattice results are compared to the HRG model curves based on the PDG2016, PDG2016+ and PDG2016+ with the inclusion of the states predicted by the hQM. The two latter spectra yield a satisfactory description of the data up to $T\simeq 145$ MeV. In the case of $\chi_4^S/\chi_2^S$, all three spectra yield a good agreement with the lattice results. Our analysis shows that the original QM overestimates these quantities because it predicts too many $|S|=2$ baryons and not enough $|S|=1$ mesons.  
In the context of future experimental measurements this study gives guidance to the RHIC, LHC and the future JLab experiments on where to focus their searches for as of yet undetected hadronic resonances.


\section*{Acknowledgements}
We acknowledge fruitful discussions with Elena Santopinto, Igor Strakovsky, Moskov Amaryan and Mark Manley.
This project was funded by the DFG grant SFB/TR55.
This material is based upon work supported by the National Science Foundation under grant no. PHY-1513864 and by the U.S. Department of Energy, Office of Science, Office of Nuclear Physics, within the framework of the Beam Energy Scan Theory (BEST) Topical Collaboration. The work of R. Bellwied is supported through DOE grant
DEFG02-07ER41521. An  award  of  computer  time  was  provided by the INCITE program.  This research used resources of the Argonne Leadership Computing Facility, which is a DOE Office of Science User Facility supported under Contract DE-AC02-06CH11357. The authors gratefully acknowledge the Gauss Centre for Supercomputing (GCS) for providing computing time for a GCS Large-Scale Project on the GCS share of the supercomputer JUQUEEN \cite{juqueen} at J\"ulich Supercomputing Centre (JSC), and at HazelHen supercomputer at HLRS, Stuttgart. The authors gratefully acknowledge the use of the Maxwell Cluster and the advanced support from the Center of Advanced Computing and Data Systems at the University of Houston.

\bibliography{biblio_strangeness}
\bibliographystyle{apsrev4-1}
\end{document}